\documentstyle[twocolumn,aps,floats]{revtex}

\begin{document}

\draft \tolerance = 10000

\setcounter{topnumber}{1}
\renewcommand{\topfraction}{0.9}
\renewcommand{\textfraction}{0.1}
\renewcommand{\floatpagefraction}{0.9}
\newcommand{\br}{{\bf r}}

\twocolumn[\hsize\textwidth\columnwidth\hsize\csname
@twocolumnfalse\endcsname

\title{ About Small Corrections to General Relativity: Gravitational
Acceleration  Depends of the Accelerated Bodies Masses (Principle of
Equivalence is Not Absolute in the Multifractal Universe) }
\author{L.Ya.Kobelev\\
 Department of  Physics, Urals State University \\ Lenina Ave., 51,
Ekaterinburg, 620083, Russia  \\ E-mail: leonid.kobelev@usu.ru} \maketitle

\begin{abstract}
In the  frame of the multifractal theory of time and space ( in this model
our universe is consisting of real time and space fields and is the
multifractal universe) in the works \cite{kob1}-\cite{kob16 } some of
problems were analyzed: how the fractional dimensions  of real fields of
time and space influence on behavior of different physical phenomena. In
this paper it is shown that in the multifractal theory of time and space
the corrections of general relativity to Newton gravitational forces are
explaining very simple in Newton approach. It is shown also that there is
dependence (though very small) of bodies gravitational acceleration from
their masses, so the principle of equivalence in multifractal universe is
not absolute and it is only a very good approach.  Thus taking into
account the multifractal dimensions of time gives small corrections to
gravitation forces of Einstein general relativity and explain it in Newton
approach. These corrections are equal to $F_{g}(r_{0}m a)(Mr_{m})^{-1}$
where $F_{g}$ is the gravitational force originated by mass $M$ at the
distance $r$ where the mass $m$ is located, $r_{0}$ is the gravitational
radius of the mass $M$ and the $r_{m}$ is the mean radius of the mass $m$,
$a$  is a numerical factor depending of distribution of masses of $m$ .
The possibility of experimental checking this effect in weak gravitational
fields is analyzed . It makes for perihelion of Mercury rotation the
correction to known results of Einstein relativity. This correction is
equal $\sim 0,5$ of percent of Einstein corrections.
\begin{center}
CONTENTS:
\end{center}
1. Introduction\\2. Newton Equations in the Multifractal Universe. \\3. Is
Gravitational Acceleration  Depends of the Mass of the Accelerated
Body?\\4. Experimental Checking\\5. Conclusions
\end{abstract}

 \vspace{1cm}

]

\section {Introduction}

The fractal model of space and time (Kobelev \cite{kob1}- \cite{kob16})
treats the time and the space with fractional dimensions as real fields.
Some information about the physical model of time and space based on these
works we use in this paper. Our universe is formed only by these fields ,
i.e. universe is fractional real material time and fractional real
material space and all other fields are born by the fields of time and
space. As the time and the space are material fields with fractional
dimensions and multifractal structure (multifractal sets,) they are
defined on the sets of their carriers of measure (physical vacuum of our
universe). In each the time (or the space) point ( "points" are approach
for very small "intervals" of time or space and "intervals" are the
multifractal sets with global dimensions for their sets, that play role of
local dimensions for universe in whole) the fractional dimensions of time
(or space) determine the densities of Lagrangians energy for all physical
fields ( or the new physical fields for space) in these points. Time and
space are binding by relation $d{\bf r}^{2}-c^{2}dt^{2}=0$ (this relation
is only a good approach, more precise relations see at \cite{kob1}-
\cite{kob2}). As the real fields time and space own by huge supply of
energy ( the question about their energy was considered partly at
\cite{kob16}) and these energies may be evaluated. The purpose of this
paper is more detailed consideration of very important problem ( in the
frame of mathematical formalism of the multifractal model of time and
space introduced in the papers of Kobelev and presented in \cite{kob1}-
\cite{kob16}): are the dependence of acceleration of any body with mass
$m$ at its mass exists or not exists (if acceleration is born by
gravitational force)?  The general relativity is based on the principle of
equivalence and deny the positive answer on this question a priori. A
posteriori this problem had not decided till now, because, as it seems,
the all experiments had fulfilled till now have small unexplained
corrections. For example, the Mercury perihelium rotation has unexplained
corrections to general relativity results that consist $0.5.$ percent of
general relativity corrections. The multifractal theory of time and space
may explain this correction and gives dependence of acceleration of body
at its mass (though very small). Thus the principle of equivalence of
general relativity lose its absolute character and become a good approach
principle. The reason of it lay in the multifractal nature of time and
space, i.e. in the multifractal nature of our universe. Time and space
build up the universe and by means of their multifractal dimensions
construct the known picture of all physical fields. So the FD of mass of
moving body have dependence of gravitational energy (see Kobelev
\cite{kob1}-\cite{kob3}) and this dependence is not be cancelled even in
Newton equation (the latter is the rude approach of scalar gravitation or
the registration only the scalar component of tensor gravitational field).
The fractional dimensions of time give the corrections to gravitational
forces such as corrections of general gravity plus additional correction
depending of accelerated bodies masses . We analyze the value of the
corrections born by this effect and consider as example the Mercury
perihelium rotation. Note that the general relativity postulating of
principle of equivalence not be rigorous in the multifractal universe and
because of the other simple reason: all systems of reference in real time
and space fields of multifractal universe are absolute systems and can be
differentiate one of another because of absolute character of time-space
fields and connection of system of reference with absolute fields. Only
the smallness of fractional corrections to topological time dimension in
our domain of universe allows to use the principle of equivalence as a
very good approach in our life and science and it works beautiful though
it is not rigorous .

\section {Newton equations in the multifractal universe}

In this paragraph we write down  the modified Newton equations in the
multifractal time universe in the presence of gravitational forces only
(see \cite{kob1}-\cite{kob3})
\begin{equation}    \label{1}
D_{-,t}^{d_{t}(r,t)}D_{+,t}^{d_{t}(r,t)} {\bf r}(t)=
D_{+,r}^{d_{r}}\Phi_{g}({\bf r}(t))
\end{equation}
\begin{equation}    \label{2}
D_{-,r}^{d_{r}}D_{+,r}^{d_{r}}\Phi_{g}({\bf r}(t))+
\frac{b_{g}^{2}}{2}\Phi_{g}({\bf r}(t))=\gamma
\end{equation}
In the Eq.(\ref{2}) the constant $b_{g}^{-1}$ has order of  size of
universe and is introduced with purpose to extend the class of functions
on which the generalized fractional derivatives concept is applicable.
These equations are not a closed system because of presence of the
fractionality of spatial dimensions. Therefore we approximate the
fractional derivatives with respect to space coordinate as $D_{+,{\bf
r}}^{d_{{\bf r}}}\approx {\bf \nabla}$, i.e. approximate them by usual
space derivatives. In the Eqs. (\ref{1})-(\ref{2}) is used the integral
functionals $D_{+,t}^{d_{t}}$ (both left-sided and right-sided) which are
suitable to describe the dynamics of functions defined on multifractal
sets (see \cite{kob1}-\cite{kob3}, \cite{kob7}). These functionals are
simple and natural generalization of the Riemann-Liouville fractional
derivatives and integrals and read:
\begin{equation} \label{3}
D_{+,t}^{d}f(t)=\left( \frac{d}{dt}\right)^{n}\int_{a}^{t}
\frac{f(t^{\prime})dt^{\prime}}{\Gamma
(n-d(t^{\prime}))(t-t^{\prime})^{d(t^{\prime})-n+1}}
\end{equation}
\begin{equation} \label{4}
D_{-,t}^{d}f(t)=(-1)^{n}\left( \frac{d}{dt}\right)
^{n}\int_{t}^{b}\frac{f(t^{\prime})dt^{\prime}}{\Gamma
(n-d(t^{\prime}))(t^{\prime}-t)^{d(t^{\prime})-n+1}}
\end{equation}
where $\Gamma(x)$ is Euler's gamma function, and $a$ and $b$ are some
constants from $[0,\infty)$. In these definitions, as usually for
Riemann-Liouville derivatives, $n=\{d\}+1$ , where $\{d\}$ is the integer
part of $d$ if $d\geq 0$ (i.e. $n-1\le d<n$) and $n=0$ for $d<0$. If
$d=const$, the generalized fractional derivatives (GFD)
(\ref{3})-(\ref{4}) coincide with the Riemann - Liouville fractional
derivatives ($d\geq 0$) or fractional integrals ($d<0$). When
$d=n+\varepsilon (t),\, \varepsilon (t)\rightarrow 0$, GFD can be
represented by means of integer derivatives and integrals. For $n=1$, i.e.
$d=1+\varepsilon$, $\left| \varepsilon \right| <<1$ it is possible to
obtain:
\begin{eqnarray}\label{5}
D_{+,t}^{1+\varepsilon }f({\mathbf r}(t),t)\approx
\frac{\partial}{\partial t} f({\mathbf r}(t),t)+ \nonumber \\ +
a\frac{\partial}{\partial t}\left[\varepsilon (r(t),t)f({\mathbf
r}(t),t)\right]+ \frac{\varepsilon ({\mathbf r}(t),t) f({\mathbf
r}(t),t)}{t}
\end{eqnarray}
where $a$ is a $constant$ and determined by  choice of the rules of
regularization of integrals (\cite{kob1}-\cite{kob2}, \cite{kob7}) (for
more detailed see \cite{kob7}) and the last additional member in the right
hand side of (\ref{5}) is very small (in this member the $t$ in the
denominator begins from "big-bang") so when the problem of irreversibility
is not researching this member may be omitted . The selection of the rule
of regularization that gives a real additives for usual derivative in
(\ref{3}) yield $a=0.5$ or $a=1$ for $d<1$ \cite{kob1}. The functions
under the integral sign in the (\ref{3})-(\ref{4}) we consider as the
generalized functions defined on the set of the finite functions or
Gelfand functions \cite{gel}. The notions of the GFD, similar to
(\ref{3})-(\ref{4}), can also be defined and for the space variables
${\mathbf r}$. The definitions of GFD (\ref{3})-(\ref{4}) need in the
connections between the fractal dimensions of time $d_{t}({\mathbf
r}(t),t)$ and the characteristics of physical fields (say, potentials
$\Phi _{i}({\mathbf r}(t),t),\,i=1,2,..)$ or densities of Lagrangians
$L_{i}$) and it was defined in cited works. Following
\cite{kob1}-\cite{kob15}, we define this connection by the relation
\begin{equation} \label{6}
d_{t}({\mathbf r}(t),t)=1+\sum_{i}\beta_{i}L_{i}(\Phi_{i} ({\mathbf
r}(t),t))
\end{equation}
where $L_{i}$ are densities of energy (Lagrangian densities) of physical
fields, $\beta_{i}$ are dimensional constants with physical dimension of
$[L_{i}]^{-1}$ (it is worth to choose $\beta _{i}^{\prime}$ in the form
$\beta _{i}^{\prime }=a^{-1}\beta _{i}$ for the sake of independence from
the regularization constant and select the $\beta=Mc^{2}$ where $M$ is the
mass of the body that born considered gravitational field). The definition
of the time as the system of subsets and definition of the FD for $d_{t}$
(see (\ref{6})) connects the value of fractional (fractal) dimension
$d_{t}(r(t),t)$  with each time instant $t$. The latter depends both on
time $t$ and coordinates ${\mathbf r}$. If $d_{t}=1$ (an absence of
physical fields) the set of time has topological dimension equal to unity.
We bound consideration only the case when relations
$d_{t}=1-\varepsilon({\mathbf r}(t),t))$, $|\varepsilon|\ll 1$ are
fulfilled.  In that case the GFD (as was shown in \cite{kob7} ) may be
represented (as a good approach) by ordinary derivatives and relation
(\ref{1}), (\ref{5})) are valid.  Now we can determine the $d_{t}$ for
distances much larger than the gravitational radius $r_{0}$ (for problem
of a body motion in the field of spherical-symmetric gravitating center)
as (see \cite{kob1} for more details)
\begin{equation} \label{7}
d_{t}\approx 1+\beta_{g}\Phi_{g}+\beta_{g} \Phi_{m}
\end{equation}
where $Phi_{g}$ is the gravitational potential born by the mass $M$ in the
center of body with mass $m$, $\Phi_{m}$ is the gravitational potential
born by the body with mass $m$ in its center. So the equations
(\ref{1})-(\ref{2}) reed ( we used for GFD approach of
(\ref{5})(\cite{kob16}))
\begin{equation}\label{8}
  [1-2\varepsilon({\bf r(t)}]\frac{d^{2}}{dt^{2}}{\bf r}= {\bf F}_{g}
\end{equation}
where
\begin{equation}\label{9}
  {\bf F}_{g}= -{\bf \nabla}\frac{\gamma M}{r} ,
  \varepsilon=\beta_{g}(\Phi_{g}+\Phi_{m})
\end{equation}
We had neglected by  the fractional parts of spatial dimensions and by the
contributions from the term with $b_{g}^{-1}$. Now we take  the
$\beta_{g}$ as $\beta_{g}=c^{-2}$ for potentials (or $\beta=(Mc)^{-2}$ for
Lagrangian density $L$). Let the body with mass $m$ moves in the
gravitational field of the body with mass $M$ on the distance $r$
($r>>r_{0}$) where $r_{0}$ is the gravitational radius of the body with
mass $M$) and let $M >>m $. Then it is possible to consider these bodies
as points masses. Now we more precisely calculate corrections to the time
dimensions in the Eq.(\ref{9}). At the point $r$, where is a center of
point mass $m$, there are two contributions in the FD of $d_{t}$: one is
caused by the field of the body with mass $M$, another is caused by the
mean gravitational field of the body with mass $m$ and with mean radius
$r_{m}$ . The latter is
\begin{eqnarray}\label{A}\nonumber
 \Phi_{m}= \frac{a m\gamma}{r_{m}}\nonumber
\end{eqnarray}
where $a$ is number factor defining by distribution of mass within body
with mass $m$. This factor is near unity ( for uniform distribution it
equal $1.5$). If we want to describe the Mercury perihelium rotation, than
it is necessary to use the law of conservation of energy . From the energy
conservation law (now this law is only approximate law ( though a very
good for $d_{t}\sim 1$) since our theory and mathematical approaches have
used apply only to open systems) obtain
\begin{eqnarray}\label{10} \nonumber
& &\left[1-\frac{2\gamma M}{c^{2}r}(1+\frac{a m r}{M r_{m}})\right]
\left(\frac{\partial r(t)}{\partial t}\right)^{2}-\frac{2mc^{2}}r +\\& &
+\left[1-\frac{2\gamma M}{c^{2}r}(1+\frac{a m r}{M r_{m}})\right]r^{2}
\left(\frac{\partial \varphi(t)}{\partial t}\right)^{2}=2E
\end{eqnarray}
Here we used the approximate relation between the generalized fractional
derivative and usual integer-order derivative (\ref{5}),(\ref{8}) and
notations corresponding to conventional description of the motion of mass
$m$ near gravitating center $M$ again. The Eq.(\ref{10}) differs from the
corresponding classical limit of the equations of general relativity by
the presence of additional terms in the first square bracket of the
left-hand part of the equation and the parentheses in the square bracket
in the right-hand part of this equation. The first term describes the
velocity alteration during gyration and is negligible in the case, when
perihelion gyration of Mercury is calculated. If we  neglect by it, then
the Eq.(\ref{10}) reduces to the corresponding classical limit of general
relativity equation with the correction born by breaking the principle of
equivalence. For large energy densities (e.g., gravitational field at
$r<r_{0}$) the Eqs.(\ref{1})-(\ref{2}) contain no divergencies \cite{kob1}
since integral-differential operators of the generalized fractional
differentiation reduced to the generalized fractional integrals (see
(\ref{1})-(\ref{4})).

 \section{Is gravitational acceleration depends of the mass of the accelerated
body?}

In this paragraph we consider the problem of limited validity of the
principle of equivalence of the general relativity. It is well known that
the principle of equivalence ( i.e. the independence of acceleration of
body with mass $m$ given by the gravitational field of it mass) is the
base of the general theory of relativity. Is this principle rigorous or
only the good approach? In the multifractal universe (our universe is
non-closed system ) all the laws and the principle that rigorous for
closed systems are not rigorous and are only the good approaches for
domain of universe where physical forces are small and so the fractional
corrections to the dimensions of time and space are small too. Now we
demonstrate an approximate character of this principle  by calculating
corrections to gravitational force in the space with fractional dimensions
of time for the case of classic limit (Newton equation, see the
(\cite{kob3}). In that case the Eq.(\ref{8}) reads $(r_{0}=\frac{2\gamma
M}{c^{2}}$ and $a=1,5$)
\begin{eqnarray}\label{11}
   [1-\frac{r_{0}}{r}(1+\frac{a m r}{M r_{m}})]\frac{d^{2}}{dt^{2}}{\bf r}=
   {\bf \nabla}\frac{\gamma M}{r}={\bf F_{g}}
\end{eqnarray}
This equation may be rewritten ( if take into account that
$r_{0}{r}^{-1}\ll1$) as
\begin{equation}\label{12}
  \frac{d^{2}}{dt^{2}}{\bf r}\approx
   {\bf F_{g}}[1+\frac{r_{0}}{r}(1+\frac{a m r}{M r_{m}})]
\end{equation}
The right-hand part of the Eq.(\ref{12}) describes the gravitation force
applied to the body placed in the central gravitational field with the
correction of general relativity and the extra correction because of
violation of principle of equivalence (the member with dependence of mass
$m$).

\section{experimental checking}

The gravitational force defined by relation (\ref{12}) may be estimated
for different physical phenomena. It gives for effect of Mercury
perihelium rotation the additional correction near $0.004a $ of
corrections of general relativity and results to full coincidence the
experimental data for this effect and theoretical calculations. We stress
that in multifractal universe there is dependence of the gravitational
acceleration from mass of accelerated body. That dependence is very small
in the considered case, but has principal role and shows the limitation of
the principle of equivalence. Such dependence is natural for the
multifractal universe, as all systems of reference in it are absolute
systems. So any equivalence of them is treated as approximation (very good
in the domains with almost topological dimensions of time and space). Let
us write the relation for relative change of part of acceleration
depending of mass the accelerating body
\begin{equation}\label{13}
\frac{a- F_{g}}{a}=\frac{r_{0}}{r}(1+\frac{3m r}{2M r_{m}})
\end{equation}
The Eq. (\ref{13}) allows to estimate the role of breaking of the
principle of equivalence for different physical phenomena. It is easy
examine that our conclusions do not contradict the experiments in this
domain and give small corrections to theoretical calculations of general
relativity .

\section{conclusions}

The main results of paper are:\\1. It is demonstrated the existence of
very small gravitational acceleration in the multifractal universe with
dependence of masses of accelerated bodies; \\2. Our result testify about
of non-absolute character of the equivalence principle of the general
relativity, but allows to use it as a very good approximated law for the
weak gravitational fields;\\3. We pay attention also that for describing
the experiments which usually are explained only by the general
relativity, in the fractal theory of time and space for for this purpose
it is enough the approach of Newton equation in the time with fractional
dimensions. Certainly, the consideration of relativistic tensor
gravitation fields gives the same results in the classic limit.

\end{document}